# Alternative Formulation for Duality-Symmetric Eleven-Dimensional Supergravity Coupled to Super M-5-Brane[1]

Hitoshi NISHINO

*Department of Physics
University of Maryland
College Park, MD 20742-4111, USA*
E-Mail: nishino@nscpmail.physics.umd.edu

## Abstract

We present an alternative formulation of duality-symmetric eleven-dimensional supergravity with both three-form and six-form gauge fields. Instead of the recently-proposed scalar auxiliary field, we use a simpler lagrangian with a non-propagating auxiliary multiplier tensor field with eight-indices. We also complete the superspace formulation in a duality-symmetric manner. An alternative super M-5-brane action coupled to this eleven-dimensional background is also presented. This formulation bypasses the usual obstruction for an invariant lagrangian for a self-dual three-form field strength, by allowing the self-duality only as a solution for field equations, but not as a necessary condition.

---

[1]This work is supported in part by NSF grant # PHY-93-41926.

# 1. Introduction

Eleven-dimensional (11D) supergravity [1] has been known for a long time to have the field content $(e_\mu{}^m, \psi_\mu, A_{\mu\nu\rho})$. In particular the significance of the three-form gauge field $A_{\mu\nu\rho}$ with its four-form field strength was elucidated, when the supermembrane formulation [2] was established with the consistent couplings to 11D supergravity. As the general construction of $p$-brane reveals [3], there may well be an alternative formulation in 11D that has the Hodge dual seven-form field strength $F_{\mu_1\cdots\mu_7}$, instead of the four-form field strength $F_{\mu\nu\rho\sigma}$. Despite of considerable efforts to formulate such 11D supergravity theory using only the seven-form field strength, the analysis in [4] indicated that there is no such a formulation possible in 11D. This obstruction is also reflected in the fact that the lagrangian in [1] has a Chern-Simons term containing not only the field strength $F_{\mu\nu\rho\sigma}$ but also the gauge field $A_{\mu\nu\rho}$ itself, preventing any duality transformations [5] into the dual field strength $F_{\mu_1\cdots\mu_7}$. However, recent development in M-theory physics [6][7][8][9][10][11][12] suggests a slightly different formulation of such a theory as the duality-symmetric limit using both the four- and seven-form field strengths at the same time, known as super M-5-brane coupled to electric and magnetic charges. Such a formulation should maintain the manifest duality-symmetry between the four and seven-form field strengths.

Recently a component formulation for such eleven-dimensional supergravity theory has been proposed [13], in which both the six-form and a three-form gauge fields are present in a duality-symmetric way. This formulation has a constraint lagrangian with an auxiliary scalar field $a(x)$, that yields the duality relation between $F_{\mu\nu\rho\sigma}$ and $F_{\mu_1\cdots\mu_7}$. This mechanism is based on the constraint lagrangian in [14] using a gradient of the scalar auxiliary field forming an unit vector $v_\mu$. Even though this formulation uses a single scalar auxiliary field $a(x)$ [13], there is much complication for the invariance confirmation of the total action due to the non-polynomial and derivative structure of $v_\mu \equiv (\partial_\mu a)[(\partial_\rho a)^2]^{-1/2}$. This complication is also reflected in the 'field-dependent' supersymmetry algebra $\{Q_\alpha, Q_\beta\} = (\gamma^m)_{\alpha\beta}[P_m - (\partial_m a)\mathcal{G}]$.[2]

In the present paper, we propose an alternative formulation of duality-symmetric supergravity theory in 11D, that has a much simpler constraint lagrangian, based on the general technique in ref. [18]. Our lagrangian has a tensor auxiliary field with eight indices, which are not totally anti-symmetric. We will show the simplicity of the confirmation of invariance of the total action in our formulation. We also re-formulate our system in superspace, that gives the confirmation of the validity of our theory. Additionally, we present a new but simple super M-5-brane action coupled to our duality-symmetric 11D supergravity backgrounds. Our action bypasses the problem with an invariant lagrangian for a self-dual three-form field strength [7], using certain constraint lagrangians.

---

[2]We point out that this algebra has a resemblance to the recent formulation in higher-dimensional supergravity/supersymmetry, F- or S-theories using null-vectors [15], multi-locality [16], or a gradient of a scalar field [17].

## 2. 11D Lagrangian in Component Formulation

Since our result is simple, it is better for us to give it first, leaving the associated remarks later. Our field content $(e_\mu{}^m, \psi_\mu, A_{\mu\nu\rho}; B_{\mu_1\cdots\mu_6}, \Lambda^{m_1\cdots m_4 n_1\cdots n_4})$ is almost the same as that of Cremmer et al. [1], except that we have an additional six-form gauge field $B_{\mu_1\cdots\mu_6}$ [3] and a tensor auxiliary field $\Lambda^{m_1\cdots m_4 n_1\cdots n_4}$. Our total action $I$ is simply a sum of three actions $I_0$, $I_1$ and $I_2$, where $I_0$ has the original Cremmer-Julia-Scherk lagrangian [1][4], and $I_1$ and $I_2$ are our deliberately chosen new actions:

$$I \equiv I_0 + I_1 + I_2 \ , \quad I_0 \equiv \int d^{11}x\, \mathcal{L}_0 \ , \quad I_1 \equiv \int d^{11}x\, \mathcal{L}_1 \ , \quad I_2 \equiv \int d^{11}x\, \mathcal{L}_2 \ , \tag{2.1}$$

$$\mathcal{L}_0 \equiv -\tfrac{1}{4} e R(\omega) - i \tfrac{1}{2}\, e\overline{\psi}_\mu \gamma^{\mu\nu\rho} D_\nu(\tfrac{\omega+\widehat{\omega}}{2})\psi_\rho - \tfrac{1}{48} e F_{[4]} F^{[4]} + \tfrac{2}{(144)^2}\epsilon^{[3][4][4]'} A_{[3]} F_{[4]} F_{[4]'}$$
$$+ \tfrac{1}{192} e[\, (\overline{\psi}_\mu \gamma^{\mu\nu\rho\sigma\tau\omega}\psi_\nu) + 12(\overline{\psi}^\rho \gamma^{\sigma\tau}\psi^\omega)\,](F_{\rho\sigma\tau\omega} + \widehat{F}_{\rho\sigma\tau\omega}) \ , \tag{2.2}$$

$$\mathcal{L}_1 \equiv +\tfrac{1}{2} e \Lambda^{m_1\cdots m_4 n_1\cdots n_4} \widehat{\mathcal{F}}_{m_1\cdots m_4} \widehat{\mathcal{F}}_{n_1\cdots n_4} \ , \tag{2.3}$$

$$\mathcal{L}_2 \equiv +\tfrac{1}{2} e \beta \widehat{\mathcal{F}}_{[4]}^2 \tag{2.4a}$$

$$= +\tfrac{1}{2}\beta e \widehat{F}_{[4]}^2 + \tfrac{1}{420}\beta e \widehat{G}_{[7]}^2 - \tfrac{1}{7!}\beta \epsilon^{[4][7]} \widehat{F}_{[4]} \widehat{G}_{[7]} \ . \tag{2.4b}$$

The symbol $[n]$ in general denotes the normalized anti-symmetric indices, e.g., $F_{[4]}F^{[4]} \equiv F_{\mu_1\cdots\mu_4} F^{\mu_1\cdots\mu_4}$, in order to save space.[5] As usual, other relevant quantities are such as

$$\widehat{F}_{\mu\nu\rho\sigma} \equiv 4\partial_{[\mu} A_{\nu\rho\sigma]} - 3(\overline{\psi}_{[\mu}\gamma_{\nu\rho}\psi_{\sigma]}) \ , \quad \widehat{\omega}_{\mu rs} \equiv \omega_{\mu rs} + \tfrac{i}{4}(\overline{\psi}_\rho \gamma_{\mu rs}{}^{\rho\sigma}\psi_\sigma) \ , \tag{2.5}$$

with the Lorentz connection $\omega_{\mu rs}$ containing $\psi$-torsion with $\gamma^{[5]}$ as well as the $\gamma^{[1]}$-matrices [1]. All the *hatted* fields are supercovariantized in component formulation [19]. The second action $I_1$ is our deliberately chosen constraint action with $\Lambda^{[4][4]'}$ as a lagrange multiplier. The $\Lambda^{[4][4]'}$ is a non-propagating multiplier field, which is *not* totally antisymmetric in all the eight indices, but instead with the (anti)symmetry

$$\Lambda^{m_1 m_2 m_3 m_4 n_1 n_2 n_3 n_4} = +\Lambda^{n_1 n_2 n_3 n_4 m_1 m_2 m_3 m_4} = -\Lambda^{m_2 m_1 m_3 m_4 n_1 n_2 n_3 n_4} \ , \quad etc. \tag{2.6}$$

Note also that all the indices in (2.3) are chosen to be local Lorentz indices, for a technical reason to be mentioned later.[6] These features will be important, when we confirm the invariance under supersymmetry. The $\widehat{\mathcal{F}}_{[4]}$ is defined by

$$\widehat{\mathcal{F}}_{m_1\cdots m_4} \equiv \widehat{F}_{m_1\cdots m_4} - \tfrac{1}{7!}\epsilon_{m_1\cdots m_4}{}^{n_1\cdots n_7} \widehat{G}_{n_1\cdots n_7} \ , \tag{2.7}$$

---

[3]We use the symbol $B$ instead of $A$ for the six-form gauge field in this paper to distinguish it from the three-form gauge field.

[4]In this section of component formulation, we use the notation [1] $(\eta_{mn}) =$ diag. $(+,-,\cdots,-)$, $\epsilon^{012\cdots 9\, 10} = +1$. We use $m, n, \cdots = (0), (1), \cdots, (10)$ for local Lorentz indices, while $\mu, \nu, \cdots = 0, 1, \cdots, 10$ for curved indices.

[5]This *normalized* $[n]$-symbol is common to all the sections. In this paper we avoid the usage of differential forms due to drawbacks, when confirming supersymmetric invariance of actions.

[6]Whenever the distinction between the local Lorentz and curved indices are crucial, we avoid the usage of the symbol $[n]$.

$$G_{\mu_1\cdots\mu_7} \equiv 7\partial_{[\mu_1} B_{\mu_2\cdots\mu_7]} - 35 F_{[\mu_1\cdots\mu_4} A_{\mu_5\mu_6\mu_7]} \quad , \tag{2.8}$$

$$\widehat{G}_{\mu_1\cdots\mu_7} \equiv G_{\mu_1\cdots\mu_7} - i\tfrac{21}{2}(\overline{\psi}_{[\mu_1}\gamma_{\mu_2\cdots\mu_6}\psi_{\mu_7]}) \quad . \tag{2.9}$$

Here the field strength $G_{[7]}$ contains the Chern-Simons form as expected from the consistency in superspace formulation [20]. The lagrangian $\mathcal{L}_1$ is similar to those constraint lagrangians in [18]. The $\beta$ in $\mathcal{L}_2$ is an arbitrary real constant subject to conditions (2.21). Eq. (2.4b) for $\mathcal{L}_2$ is to make the $G_{[7]}^2$-kinetic term explicit. Note that the last $\epsilon\widehat{F}\widehat{G}$-term in (2.4b) is a total divergence at the lowest order, so it should be regarded as a trilinear term. It is helpful to remember that $\mathcal{L}_2$ can be easily obtained by the simple field redefinition of $\Lambda_{[4]}{}^{[4]'}$ by a product of Kronecker's delta (Cf. (2.15) below).

Our supersymmetry transformation rule is

$$\delta_Q e_\mu{}^m = -i(\overline{\epsilon}\gamma^m\psi_\mu) \quad , \tag{2.10a}$$

$$\begin{aligned}\delta_Q \psi_\mu =& +D_\mu(\widehat{\omega})\epsilon + i\tfrac{1}{144}(\gamma_\mu{}^{[4]}\widehat{F}_{[4]} - 8\gamma^{[3]}\widehat{F}_{\mu[3]})\epsilon \\ & - i\tfrac{1}{6}(\gamma_{\mu[4]}\epsilon\,\Lambda^{[4][4]'} - 8\gamma^{[3]}\epsilon\,\Lambda_{\mu[3]}{}^{[4]'})\widehat{\mathcal{F}}_{[4]'} \\ & - i\tfrac{1}{6}\beta(\gamma_\mu{}^{[4]}\epsilon\,\widehat{\mathcal{F}}_{[4]} - 8\gamma^{[3]}\epsilon\,\widehat{\mathcal{F}}_{\mu[3]}) \equiv \widehat{\mathcal{D}}_\mu\epsilon \quad , \end{aligned}\tag{2.10b}$$

$$\delta_Q A_{\mu\nu\rho} = +\tfrac{3}{2}(\overline{\epsilon}\gamma_{[\mu\nu}\psi_{\rho]}) \quad , \tag{2.10c}$$

$$\delta_Q B_{\mu_1\cdots\mu_6} = +3i(\overline{\epsilon}\gamma_{[\mu_1\cdots\mu_5}\psi_{\mu_6]}) - 20 A_{[\mu_1\mu_2\mu_3}(\delta_Q A_{\mu_4\mu_5\mu_6]}) \quad , \tag{2.10d}$$

$$\begin{aligned}\delta_Q \Lambda^{m_1\cdots m_4 n_1\cdots n_4} =& \Big[+i\tfrac{3}{2}(\overline{\epsilon}\gamma^\mu\psi_\mu)\Lambda^{m_1\cdots m_4 n_1\cdots n_4} + 4i(\overline{\epsilon}\gamma_r\psi^{[m_1})\Lambda^{m_2 m_3 m_4]rn_1\cdots n_4} \\ & - 24i(\overline{\epsilon}\gamma^{\rho[2]}\psi^\sigma)\Lambda_{\rho\sigma[2]'}{}^{m_1\cdots m_4}\Lambda_{[2]}{}^{[2]'n_1\cdots n_4} - 96i(\overline{\epsilon}\gamma_\rho\psi_\sigma)\Lambda_{\rho[3]}{}^{m_1\cdots m_4}\Lambda_\sigma{}^{[3]n_1\cdots n_4} \\ & + i\tfrac{1}{6}(\overline{\epsilon}\gamma^{[4][4]'\nu}\psi_\nu)\Lambda_{[4]}{}^{m_1\cdots m_4}\Lambda_{[4]'}{}^{n_1\cdots n_4} - 24i(\overline{\epsilon}\gamma^{\rho[2][2]'}\psi_\rho)\Lambda_{[2]}{}^{[2]''m_1\cdots m_4}\Lambda_{[2]'[2]''}{}^{n_1\cdots n_4} \\ & + 32i(\overline{\epsilon}\gamma_{[2]}{}^{[3]}\psi_\rho)\Lambda^{\rho\sigma[2]m_1\cdots m_4}\Lambda_{\sigma[3]}{}^{n_1\cdots n_4} + 12i(\overline{\epsilon}\gamma^\mu\psi_\mu)\Lambda_{[4]}{}^{m_1\cdots m_4}\Lambda^{[4]n_1\cdots n_4}\Big]\Big|_{\Lambda\to\Lambda+\delta\cdots\delta} \\ & + (m_i\leftrightarrow n_i) \quad . \end{aligned}\tag{2.10e}$$

The last manipulation $(m_i\leftrightarrow n_i)$ is needed to make the r.h.s. to have the same symmetry as $\Lambda^{[4][4]'}$, while the operation $|_{\Lambda\to\Lambda+\delta\cdots\delta}$ is for the replacement $\Lambda_{m_1\cdots m_4}{}^{n_1\cdots n_4} \to \Lambda_{m_1\cdots m_4}{}^{n_1\cdots n_4} + \beta\delta_{[m_1}{}^{[n_1}\cdots\delta_{m_4]}{}^{n_4]}$ of all the $\Lambda_{[4]}{}^{[4]'}$ in the square bracket. The significance of this operation will be clarified shortly. The rules (2.10c) and (2.10d) justify the coefficients of $\psi$-dependent terms in the supercovariant field strengths given above. Note also that all the indices in (2.10e) are local Lorentz indices, but *not* curved indices, because the difference will yield gravitino-linear terms out of elfbein variations.

Before confirming the invariance of our action under supersymmetry, we first consider the field equations of all the fields. The $\Lambda^{[4][4]'}$-field equation immediately gives the duality constraint

$$\widehat{\mathcal{F}}_{m_1\cdots m_4} \equiv \widehat{F}_{m_1\cdots m_4} - \tfrac{1}{7!}\epsilon_{m_1\cdots m_4}{}^{n_1\cdots n_7}\widehat{G}_{n_1\cdots n_7} = 0 \quad , \tag{2.11}$$

namely the duality between $\widehat{F}_{[4]}$ and $\widehat{G}_{[7]}$ like [13]. This is because the indices $[4]$ and $[4]'$ in the original $\Lambda$-field equation $\widehat{\mathcal{F}}_{[4]}\widehat{\mathcal{F}}_{[4]'} = 0$ are free independent indices, implying that $\widehat{\mathcal{F}}_{[4]} = 0$. Once eq. (2.11) is satisfied, we immediately see that the contribution of both $I_1$ and $I_2$ to the $A_{[3]}$-field equation vanishes, because it contains one factor of $\widehat{\mathcal{F}}_{[4]} = 0$ by (2.11):

$$\widehat{D}_\mu \widehat{F}^\mu{}_{\rho\sigma\tau} - \tfrac{1}{576} e^{-1} \epsilon_{\rho\sigma\tau}{}^{[4][4]'} \widehat{F}_{[4]} \widehat{F}_{[4]'} = 0 \ . \tag{2.12}$$

The same is also true for the gravitino field equation, even though $\psi_\mu$ is involved in $\widehat{\mathcal{F}}_{[4]}$:

$$i\gamma^{\mu\nu\rho} \widehat{\mathcal{R}}_{\nu\rho} = 0 \ , \tag{2.13}$$

where $\widehat{\mathcal{R}}_{\mu\nu} \equiv \widehat{\mathcal{D}}_\mu \psi_\nu - \widehat{\mathcal{D}}_\nu \psi_\mu$ is the supercovariant field strength of the gravitino in our notation. Here the meaning of *hat* is the same one in the r.h.s. in (2.10b). An important feature here is that this gravitino field equation has not only the contribution from $\mathcal{L}_0$, but also that from $\mathcal{L}_1$ and $\mathcal{L}_2$ through the supercovariantized $\widehat{G}_{[7]}$, proportional to $\Lambda$ and $\widehat{\mathcal{F}}$. This can be confirmed explicitly by taking the variation of $\widehat{\mathcal{F}}_{[7]}$ with respect to the gravitino. The $B_{[6]}$-field equation is automatically satisfied by (2.11), due to the presence of $B_{[6]}$ only in $\widehat{\mathcal{F}}_{[4]}$. Now we see that the elfbein field equation is not affected, because of the bilinear structure of $\mathcal{L}_1$ in $\widehat{\mathcal{F}}$, as other field equations:

$$\widehat{R}_{\mu\nu} - \tfrac{1}{3}(\widehat{F}_{\mu[3]} \widehat{F}_\nu{}^{[3]} - \tfrac{1}{24} g_{\mu\nu} \widehat{F}_{[4]}^2) = 0 \ . \tag{2.14}$$

We therefore conclude that the only new effect of $I_1$ and $I_2$ on field equations is simply the duality equation (2.11), and none of the field equations of the original fields $e_\mu{}^m$, $\psi_\mu$, $A_{[3]}$ are affected. Moreover, the $\Lambda$-field is completely decoupled from any field equations in our system. An important consequence of the duality condition (2.11) is that the $B_{[6]}$-field has got dynamical freedom, because of its non-vanishing divergence due to this duality. However, note also that the superficial kinetic term for $B_{[6]}$ in $\mathcal{L}_2$ does *not* contribute to the $B_{[6]}$-field equation, because there is always the on-shell vanishing factor accompanying the variation. Eventually the degrees of freedom for all of these antisymmetric tensor field stay the same as $(84+84)/2 = 84$. This feature of degrees of freedom is the same as in [13].

We now confirm the invariance of our total action $I$ under supersymmetry dictated by (2.10). As a universal notation, we distinguish the Cremmer *et al.*'s supersymmetry transformation rule $\delta_Q^{(0)}$ [1] from our modified terms $\delta_Q^{(1)}$ in the transformation rule in (2.10) which is $\Lambda$-dependent or $\beta$-dependent: $\delta_Q = \delta_Q^{(0)} + \delta_Q^{(1)}$, *except* (2.10e), for which $\delta_Q \Lambda^{[4][4]'} = \delta_Q^{(0)} \Lambda^{[4][4]'}$, just for convenience of manipulations. In other words, $\delta_Q^{(1)}$ denotes all the $\Lambda$-dependent and $\beta$-dependent terms in (2.10). We next introduce a technique that drastically simplifies the whole computation. Note that $\mathcal{L}_2$ can be simply obtained from $\mathcal{L}_1$ by the field redefinition

$$\Lambda_{m_1 \cdots m_4}{}^{n_1 \cdots n_4} \to \Lambda_{m_1 \cdots m_4}{}^{n_1 \cdots n_4} + \beta \delta_{[m_1}{}^{[n_1} \cdots \delta_{m_4]}{}^{n_4]} \ . \tag{2.15}$$

This implies that to perform the above invariance check, we do not have to consider $\mathcal{L}_2$ as an independent lagrangian, but once the invariance of $I_0 + I_1$ is confirmed for the transformation rule (2.10) with $\beta = 0$, then we can extrapolate this result to the general case $\beta \neq 0$, just by the field redefinition (2.15).

Recalling now $\delta_Q^{(0)} \mathcal{L}_0 = 0$ [1], we can understand the invariance of our action under supersymmetry when $\beta = 0$ as

$$\begin{aligned}
0 &\stackrel{?}{=} \delta_Q(\mathcal{L}_0 + \mathcal{L}_1) = (\delta_Q^{(0)} + \delta_Q^{(1)})(\mathcal{L}_0 + \mathcal{L}_1) \\
&= \delta_Q^{(0)} \mathcal{L}_0 + \delta_Q^{(1)}(\mathcal{L}_0 + \mathcal{L}_1) + \delta_Q^{(0)} \mathcal{L}_1 \\
&= e(\delta_Q^{(1)}\overline{\psi}_\mu)\widehat{\mathcal{S}}^\mu + \delta_Q^{(0)} \mathcal{L}_1 \\
&= +i\tfrac{1}{6} e\Big(\overline{\epsilon}\gamma_\mu{}^{[4]}\Lambda_{[4]}{}^{[4]'}\widehat{\mathcal{F}}_{[4]'} + 8\overline{\epsilon}\gamma^{(3)}\Lambda_{\mu[3]}{}^{[4]}\widehat{\mathcal{F}}_{[4]}\Big)(\widehat{\mathcal{S}}^{\mu(0)} + \widehat{\mathcal{S}}^{\mu(1)}) \\
&\quad + \tfrac{1}{2}[\delta_Q^{(0)}(e\Lambda^{m_1\cdots m_4 n_1\cdots n_4})]\widehat{\mathcal{F}}_{m_1\cdots m_4}\widehat{\mathcal{F}}_{n_1\cdots n_4} + e\Lambda^{m_1\cdots m_4 n_1\cdots n_4}(\delta_Q^{(0)}\widehat{\mathcal{F}}_{m_1\cdots m_4})\widehat{\mathcal{F}}_{n_1\cdots n_4} \ . \quad (2.16)
\end{aligned}$$

Here $\widehat{\mathcal{S}}^\mu$ is for the l.h.s. of the gravitino field equation, while $\widehat{\mathcal{S}}^{\mu(1)}$ denotes the terms in $\widehat{\mathcal{S}}^\mu$ coming only from $I_1$, as is easily computed:

$$\widehat{\mathcal{S}}^\mu \equiv e^{-1}\frac{\delta}{\delta\psi_\mu}(\mathcal{L}_0 + \mathcal{L}_1) = -i\tfrac{1}{2}\gamma^{\mu\nu\rho}\widehat{\mathcal{R}}_{\nu\rho} \equiv \widehat{\mathcal{S}}^{\mu(0)} + \widehat{\mathcal{S}}^{\mu(1)} \ , \quad (2.17)$$

$$\widehat{\mathcal{S}}^{\mu(1)} \equiv e^{-1}\frac{\delta}{\delta\psi_\mu}\mathcal{L}_1 = -6\gamma_{[2]}\psi_\nu\Lambda_\mu{}^{\nu[2][4]}\widehat{\mathcal{F}}_{[4]} - \tfrac{1}{2}\gamma^{\mu\nu[4]}\psi_\nu\Lambda_{[4]}{}^{[4]'}\widehat{\mathcal{F}}_{[4]'} \ . \quad (2.18)$$

We now easily see that all the terms linear in $\widehat{\mathcal{S}}^{\mu(0)}$ cancel themselves by the aid of the identities such as

$$\begin{aligned}
\delta_Q^{(0)}\widehat{\mathcal{F}}_{mnrs} &= -i\tfrac{1}{6}(\overline{\epsilon}\gamma_{mnrs}{}^t\widehat{\mathcal{S}}_t^{(0)}) + i\tfrac{4}{3}(\overline{\epsilon}\gamma_{[mnr}\widehat{\mathcal{S}}_{s]}^{(0)}) \\
&\quad - i(\overline{\epsilon}\gamma^\mu\psi_\mu)\widehat{\mathcal{F}}_{mnrs} + 4i(\overline{\epsilon}\gamma_{[m|}\psi^t)\widehat{\mathcal{F}}_{t|nrs]} \ , \quad &(2.19\text{a}) \\
\delta_Q^{(0)}\widehat{F}_{mnrs} &= +3(\overline{\epsilon}\gamma_{[mn}\widehat{\mathcal{R}}_{rs]}^{(0)}) \ , \quad &(2.19\text{b}) \\
\delta_Q^{(0)}\widehat{G}_{m_1\cdots m_7} &= -i\tfrac{21}{2}(\overline{\epsilon}\gamma_{[m_1\cdots m_5}\widehat{\mathcal{R}}_{m_6 m_7]}^{(0)}) + 7i(\overline{\epsilon}\gamma_n\psi_{[m_1})\widehat{\mathcal{G}}_{m_2\cdots m_7]}{}^n \ , \quad &(2.19\text{c}) \\
\widehat{\mathcal{G}}_{m_1\cdots m_7} &\equiv +\tfrac{1}{4!}\epsilon_{m_1\cdots m_7}{}^{n_1\cdots n_4}\widehat{\mathcal{F}}_{n_1\cdots n_4} \ . \quad &(2.19\text{d})
\end{aligned}$$

We next see that the remaining $\Lambda$-dependent terms both of the types $\approx \psi\Lambda\widehat{\mathcal{F}}$ and $\psi\Lambda^2\widehat{\mathcal{F}}^2$ cancel themselves, and therefore we establish the supersymmetric invariance: $\delta_Q(I_0 + I_1) = 0$, when $\beta = 0$. As has been mentioned, since the case of $\beta \neq 0$ with $I_2$ can be re-obtained by the simple field redefinition (2.15), the total action $I_0 + I_1 + I_2$ is also invariant under the transformation rule (2.10) now with $\beta \neq 0$.

The advantage of the algorithm in (2.15) for the invariance check is that we have to take only the variation $\delta_Q^{(0)}$ by Cremmer et al. [1]. This considerably simplifies the computation, enabling us to fix the transformation rule for $\Lambda^{[4][4]'}$. The presence of the $\psi$-linear term in (2.19a,c) seems to be from the fact that the original Cremmer et al.'s transformation rule $\delta_Q^{(0)}\psi_\mu$ in (2.10b) is not duality-symmetric.

We next consider the on-shell closure of the gauge algebra. The only difference of our transformation rule from that in [1] is the presence of $\Lambda\widehat{\mathcal{F}}$ or $\beta\widehat{\mathcal{F}}$-terms in (2.10b), $\Lambda$-terms in (2.10e), and (2.10d) itself. This does not pose any problem for the closure of gauge algebra, seen as follows. First, the on-shell closure on $e_\mu{}^m$ stays the same, because of $\widehat{\mathcal{F}}_{[4]} = 0$. The same is also true for the closure on $A_{[3]}$ and on $B_{[6]}$. The least trivial one is the closure on $\psi_\mu$, which eventually has no problem, because of the key relation $\delta_Q \widehat{\mathcal{F}}_{mnrs} = \delta_Q^{(0)} \widehat{\mathcal{F}}_{mnrs}$ equivalent to (2.19a) on-shell. Finally the closure on $\Lambda^{m_1 \cdots m_4 n_1 \cdots n_4}$ itself has no problem, even though it looks awfully complicated, because this auxiliary field is decoupled from any field equations, and in fact, it can be gauged away by an extra symmetry mentioned next.

The absence of $\Lambda^{[4][4]'}$ from all the field equations suggests that this multiplier field might be gauged away as a non-physical field. As a matter of fact, we can see the existence of an extra symmetry [18] for $\Lambda^{[4][4]'}$:

$$\delta_\eta \Lambda_{\mu_1 \cdots \mu_4}{}^{\nu_1 \cdots \nu_4} = \Big[ + \partial_{[\mu_1} \eta_{\mu_2 \mu_3 \mu_4]}{}^{\nu_1 \cdots \nu_4} - 2 \delta_{[\mu_1|}{}^{[\nu_1|} \partial_\rho \eta_{|\mu_2 \mu_3 \mu_4]}{}^{\rho|\nu_2 \nu_3 \nu_4]} \Big] + (\mu_i \leftrightarrow \nu_i) \ , \qquad (2.20a)$$

$$\delta_\eta A_{\mu\nu\rho} = -\tfrac{6}{24\beta - 1} \eta_{\mu\nu\rho}{}^{[4]} \widehat{\mathcal{F}}_{[4]} \ , \qquad \delta_\eta B_{\mu_1 \cdots \mu_6} = -\tfrac{1}{24\beta} \epsilon_{\nu\mu_1 \cdots \mu_6}{}^{[4]} \eta_{[3][4]} \widehat{\mathcal{F}}^{[3]\nu} \ , \qquad (2.20b)$$

up to the next order terms.[7] Here $\eta_{\mu_1 \mu_2 \mu_3}{}^{\nu_1 \cdots \nu_4}$ is an arbitrary space-time dependent parameter, anti-symmetric under $[\mu_1 \mu_2 \mu_3]$ and under $[\nu_1 \cdots \nu_4]$, but with no other (anti)symmetries. Since we are interested only in the lowest-order, we are using the curved indices here. The last operation in (2.20a) is just to make the r.h.s. have the same symmetry as the l.h.s. As the examples in [18], the extra transformations for $A_{[3]}$ and $B_{[6]}$ vanish on-shell, and more importantly, this symmetry can gauge away the auxiliary field $\Lambda^{[4][4]'}$, when

$$\beta \neq 0 \ , \qquad \beta \neq \tfrac{1}{24} \ . \qquad (2.21)$$

Eventually, the only important role played by $\Lambda^{[4][4]'}$ is to yield the constraint (2.11) as a multiplier field. The special case $\beta = 0$ or $\beta = (24)^{-1}$ can be understood as a singular case, when the kinetic term for $A_{[3]}$ or $B_{[6]}$ disappears. To see this more explicitly, consider now the bosonic terms depending only $A_{[3]}$ and $B_{[6]}$ in the total lagrangian

$$\mathcal{L}_{A,B} = +\tfrac{1}{2}(\beta - \tfrac{1}{24}) F_{[4]}^2 + \tfrac{1}{420}\beta G_{[7]}^2 + \tfrac{1}{144}(\beta + \tfrac{1}{72}) \epsilon^{[3][4][4]'} A_{[3]} F_{[4]} F_{[4]'} \ , \qquad (2.22)$$

up to non-essential higher-order terms. The singular case $\beta = (24)^{-1}$ corresponds to the absence of the kinetic term of $A_{[3]}$, while the other singular case $\beta = 0$ corresponds to the absence of $G_{[7]}^2$-term. These singular cases $\beta = 0, \beta = (24)^{-1}$ do not accommodate the extra symmetry (2.20), and therefore the $\Lambda^{[4][4]'}$-field can not be gauged away as in [18]. In other words, the presence of both the $F_{[4]}^2$ and $G_{[7]}^2$-terms seem to be crucial for our formulation using the multiplier field, even though these singular cases would give the simplest lagrangians.

---

[7] Even though this expression is only at the lowest order, other higher-order terms can be also fixed which are skipped in this paper.

## 3. Superspace Formulation

Once the component formulation has been established, we are ready to consider the corresponding superspace formulation, as has been almost always the case with supergravity theories. Due to the newly introduced six-form gauge field $B_{[6]}$, we need to consider the three independent superspace Bianchi identities (BIs)

$$\nabla_{[A}T_{BC)}{}^E - T_{[AB|}{}^E T_{E|C)}{}^D - \tfrac{1}{2}R_{[AB|e}{}^f(\mathcal{M}_f{}^e)_{|C)}{}^D \equiv 0 \ , \tag{3.1a}$$

$$\tfrac{1}{4!}\nabla_{[A_1}F_{A_2\cdots A_5)} - \tfrac{1}{3!\cdot 2}T_{[A_1 A_2|}{}^B F_{B|A_3 A_4 A_5)} \equiv 0 \ , \tag{3.1b}$$

$$\tfrac{1}{7!}\nabla_{[A_1}G_{A_2\cdots A_8)} - \tfrac{1}{6!\cdot 2}T_{[A_1 A_2|}{}^B G_{B|A_3\cdots A_8)} + \tfrac{1}{(4!)^2}F_{[A_1\cdots A_4}F_{A_5\cdots A_8)} \equiv 0 \ . \tag{3.1c}$$

We call these Bianchi identities respectively $(ABC, D)$, $(A_1\cdots A_5)$ and $(A_1\cdots A_8)$-types. In this section of superspace (and the next section as well), we use the indices $A$, $B$, $\cdots$ for the local Lorentz indices in superspace, which can be either bosonic $a$, $b$, $\cdots$ or fermionic $\alpha$, $\beta$, $\cdots$. The antisymmetrization symbol in (3.1) is different from the previous section for components, because now we have e.g., $C_{[AB)} \equiv C_{AB} \pm C_{BA}$ with no factor of $1/2$. This notation is common to sections 3 and 4. As has been also known [13][20], the presence of the Chern-Simons form in (3.1c) is crucial for the lowest engineering dimensional BI at $d = 0$ via (3.4) below, also consistent with (2.8).

From the component result, we can see the relevant superspace constraints are

$$T_{\alpha\beta}{}^c = +i(\gamma^c)_{\alpha\beta} \ , \quad F_{\alpha\beta cd} = +\tfrac{1}{2}(\gamma_{cd})_{\alpha\beta} \ , \tag{3.2a}$$

$$G_{\alpha\beta c_1\cdots c_5} = -i\tfrac{1}{2}(\gamma_{c_1\cdots c_5})_{\alpha\beta} \ , \tag{3.2b}$$

$$\begin{aligned}T_{ab}{}^\gamma &= +i\tfrac{1}{144}(\gamma_b{}^{[4]}F_{[4]} + 8\gamma^{[3]}F_{b[3]})_\alpha{}^\gamma \\ &\quad - i\tfrac{1}{6}(\gamma_b{}^{[4]}\Lambda_{[4]}{}^{[4]'} + 8\gamma^{[3]}\Lambda_{b[3]}{}^{[4]'})_\alpha{}^\gamma \mathcal{F}_{[4]'} \\ &\quad - i\tfrac{1}{6}\beta(\gamma_b{}^{[4]}\mathcal{F}_{[4]} + 8\gamma^{[3]}\mathcal{F}_{b[3]})_\alpha{}^\gamma \ , \end{aligned} \tag{3.2c}$$

$$\mathcal{F}_{[4]} = F_{[4]} - \tfrac{1}{7!}\epsilon_{[4]}{}^{[7]}G_{[7]} = 0 \ . \tag{3.2d}$$

As usual, we do not put any *hat* on field strengths in superspace, due to their manifest supercovariance [21]. As is often with superspace for higher-dimensional supergravity, all the equations are essentially on-shell [21]. Since dimension $d = 2$ BI will yield the field equation $\mathcal{F}_{[4]} = 0$, as will be seen, the presence of the $\mathcal{F}$-terms in (3.2c) should not matter. However, inclusion of them is useful to re-confirm important relationships used in component formulation.

We now analyze these BIs at each engineering dimension. First of all, the BIs (3.1a) and (3.1b) are not affected, except for the $\mathcal{F}$-dependent terms in $T_{ab}{}^\gamma$ which we keep as manifest, even though they vanish on-shell. Relevantly, the $(\alpha\beta cde)$-type BI at $d = 1$ yields the field equation $\mathcal{F}_{[4]} = 0$. The $(\alpha bcde)$-type BI at $d = 3/2$ yields one of the important relationships:

$$\nabla_\alpha F_{bcde} = -\tfrac{1}{8}(\gamma_{[bc|})_{\alpha\beta}T_{|de]}{}^\beta \ , \tag{3.3}$$

which is on-shell equivalent to (2.19a) in component. All other equations out of BIs in (3.1a) and (3.1b) are formally equivalent to the case of [22][1]. The only non-trivial confirmation is for (3.1c). At $d=0$ for the $(\alpha\beta\gamma\delta c_1\cdots c_4)$-type BI, the following crucial identity is used:

$$(\gamma^e)_{(\alpha\beta|}(\gamma_{eabcd})_{|\gamma\delta)} \equiv +\tfrac{1}{8}(\gamma_{[ab|})_{(\alpha\beta|}(\gamma_{|cd]})_{\gamma\delta)} \quad, \tag{3.4}$$

which is confirmed by another identity $(\gamma^{ab})_{(\alpha\beta|}(\gamma_b)_{|\gamma\delta)} \equiv 0$. It is this identity that requires the presence of the Chern-Simons form in the field strength $G_{[7]}$ [13][20]. The next non-trivial confirmation is at $d=1$ for $(\alpha\beta c_1\cdots c_6)$-type BI, which consists of three structures of $\gamma$-matrices: (i) $\gamma_{c_1\cdots c_6}{}^{[4]}F_{[4]}$, (ii) $\gamma_{[c_1\cdots c_4|}{}^{[2]}F_{|c_5 c_6][2]}$, (iii) $\gamma_{[c_1 c_2|}F_{|c_3\cdots c_6]}$, after converting $F_{[4]}$ into $G_{[7]}$ by (3.2d). Fortunately, all of these sectors vanish by cancellation of the like terms by themselves, after the appropriate use of $\gamma$-matrix identities, such as $\gamma^{[10]} \equiv +i\epsilon^{[10]m}\gamma_m$. At $d=3/2$, we get

$$\nabla_\alpha G_{b_1\cdots b_7} = -i\tfrac{1}{480}(\gamma_{[b_1\cdots b_5|})_{\alpha\beta}T_{|b_6 b_7]}{}^\beta \quad, \tag{3.5}$$

which is easily shown to be on-shell equivalent to (2.19c), and consistent with the duality relation (2.11) or (3.2d). As usual at $d=2$, we see that the appearance of $\theta=0$ sector is consistent with the component field strength $G_{[7]}$ (2.8) with the Chern-Simons form.

## 4. Couplings to Super M-5-Brane

In this section, we try to couple our 11D supergravity background to super M-5-brane. Our action is in a sense simpler than those in [11] or [9][10][14], and circumvents the usual problem [7] for an invariant lagrangian for self-dual field strength.

Our fundamental fields in 6D are $(Z^M, g_{ij}, a_{ij}, b_{[5]}, \lambda^{ijk}_{(+)}, \lambda, \rho^{ijk}_{(+)}, \mu^{(-)(-)}_{ijklmn}, \nu^{(-)}_{ijk}).^8$ Here $Z^M$ is the 11D superspace coordinates, $a_{ij}$ is antisymmetric field, $g_{ij}$ is the 6D metric, while the auxiliary tensor density $\lambda^{ijk}_{(+)}$ and tensor $\rho^{ijk}_{(+)}$ are self-dual with respect to the indices ${}^{ijk}$. The $\mu^{(-)(-)}_{ijklmn}$ is a tensor auxiliary field, and is anti-self-dual with respect to the first three indices ${}_{ijk}$, as well as to the last three indices ${}_{lmn}$. The $\nu^{(-)}_{ijk}$ is a anti-self-dual tensor density auxiliary field.

Our total action has five parts: the first term $S_{\text{det}}$ with a determinant, the constraint terms $S_{\lambda f}$, $S_{\lambda g}$, $S_{\mu\rho\lambda}$ and $S_{\nu\rho\lambda}$:

$$S \equiv S_{\text{det}} + S_{\lambda f} + S_{\lambda g} + S_{\mu\rho\lambda} + S_{\nu\rho\lambda} \equiv \int d^6\sigma \mathcal{L} \quad, \tag{4.1}$$

$$S_{\text{det}} \equiv \int d^6\sigma \left[ \det\left(g_{ij} - \Pi_i{}^a \Pi_{ja}\right) \right]^{1/2} \equiv \int d^6\sigma \mathcal{L}_{\text{det}} \quad, \tag{4.2}$$

$$S_{\lambda f} \equiv \int d^6\sigma \lambda^{ijk}_{(+)} f_{ijk} \equiv \int d^6\sigma \mathcal{L}_{\lambda f} \quad, \tag{4.3}$$

---

[8]Our 6D notation is $(\eta_{(i)(j)}) = \text{diag.}(+,-,-,-,-,-)$, $\epsilon^{012\cdots 5} = +1$. We use $(i), (j), \cdots = (0), (1), \cdots, (5)$ for local Lorentz indices, while $i, j, \cdots = 0, 1, \cdots, 5$ for curved indices.

$$S_{\lambda g} \equiv \int d^6\sigma \left( \tfrac{2}{6!}\epsilon^{i_1\cdots i_6}\lambda g_{i_1\cdots i_6} + e\lambda \right) \equiv \int d^6\sigma \, \mathcal{L}_{\lambda g} \ , \tag{4.4}$$

$$S_{\mu\rho\lambda} \equiv \int d^6\sigma \, \mu^{(-)(-)}_{ijklmn}\rho^{ijk}_{(+)}\lambda^{lmn}_{(+)} \equiv \int d^6\sigma \, \mathcal{L}_{\mu\rho\lambda} \ , \tag{4.5}$$

$$S_{\nu\rho\lambda} \equiv \int d^6\sigma \, \nu^{(-)}_{ijk}\rho^{ijk}_{(+)}\lambda \equiv \int d^6\sigma \, \mathcal{L}_{\nu\rho\lambda} \ . \tag{4.6}$$

The field strengths $f_{[3]}$ and $g_{[6]}$ are defined by [23][24][9][11]

$$f_{ijk} \equiv \tfrac{1}{2}\partial_{[i}a_{jk]} - A_{ijk} \ , \qquad A_{ijk} \equiv \Pi_k{}^C \Pi_j{}^B \Pi_i{}^A A_{ABC} \ , \tag{4.7a}$$

$$g_{i_1\cdots i_6} \equiv \tfrac{1}{5!}\partial_{[i_1}b_{i_2\cdots i_6]} - B_{i_1\cdots i_6} + \tfrac{1}{48}a_{[i_1 i_2}F_{i_3\cdots i_6]} \ . \tag{4.7b}$$

The $\Pi_i{}^A \equiv (\partial_i Z^M) E_M{}^A(Z)$ are the pull-backs from 11D superspace to the 6D world-supervolume. As usual convention as in (4.7a), the 11D superspace indices $A, B, \cdots$ can be replaced by the 6D indices $i, j, \cdots$ by the use of the pull-backs $\Pi_i{}^A$. The 6D field strengths $f$ and $g$ contain the 11D superfield potentials like the D-brane couplings [11][23][24]. Our lagrangian $\mathcal{L}_{\mu\rho\lambda}$ or $\mathcal{L}_{\nu\rho\lambda}$ resembles those in [18], because this lagrangian is also bilinear, but is a product of different fields $\rho$ and $\lambda$. Note that the $\lambda^{[3]}$ and $\nu_{[3]}$-fields are tensor densities by definition, and $f_{ijk}$ in (4.3) needs no self-duality projector. Hence no $g_{ij}$ is involved in (4.3) and (4.4), *except* the $e\lambda$-term in the latter.

Our fermionic symmetry is dictated by the transformation rule

$$\delta_\zeta E^\alpha \equiv (\delta_\zeta Z^M) E_M{}^\alpha = \kappa_-^\alpha \equiv i(\gamma_{[3]}\zeta_+)^\alpha \rho^{[3]}_{(+)} \ , \tag{4.8a}$$

$$\delta_\zeta E^a \equiv (\delta_\zeta Z^M) E_M{}^a = 0 \ , \quad \delta_\zeta \lambda^{ijk}_{(+)} = 0 \ , \quad \delta_\zeta \rho^{ijk}_{(+)} = 0 \ , \quad \delta_\zeta \lambda = 0 \ , \tag{4.8b}$$

$$\delta_\zeta g_{ij} = \delta_\zeta(\Pi_i{}^a \Pi_{ja}) = -i(\overline{\kappa}_- \gamma_{(i}\Pi_{j)}) \ , \quad \delta_\zeta e = -i(\overline{\kappa}_- \gamma^i \Pi_i) \ , \tag{4.8c}$$

$$\delta_\zeta a_{ij} = (\delta_\zeta E^B) A_{Bij} \ , \quad \delta_\zeta b_{i_1\cdots i_5} = (\delta_\zeta E^B) B_{Bi_1\cdots i_5} \ , \tag{4.8d}$$

$$\delta_\zeta \mu^{(-)(-)}_{ijklmn} = -i\tfrac{1}{4}(\overline{\zeta}_+ \gamma_{ijk}\gamma^p \gamma_{lmn}\Pi_p) \ , \tag{4.8e}$$

$$\delta_\zeta \nu^{(-)}_{ijk} = -i\tfrac{1}{6}e(\zeta_+ \gamma_{ijk}\gamma^{lmnp}\Pi_l)f_{mnp} \ . \tag{4.8g}$$

Here $e \equiv \det(e_i{}^{(j)}) \equiv \sqrt{-g}$ is the determinant of sechsbein in 6D. The $\gamma$-matrices such as $\gamma_i$ is defined by $\gamma_i \equiv \Pi_i{}^a \gamma_a$, satisfying the 6D Clifford algebra

$$\{\gamma_i, \gamma_j\} = 2g_{ij}I \ , \tag{4.9}$$

under the embedding condition (4.17) below. We can also specify the chirality in 6D, defining $\gamma_7 \equiv \gamma^{(0)\cdots(5)}$. The $\zeta_+$ is the parameter for our fermionic symmetry, and only its positive chirality part is involved in our transformation. This fermionic symmetry deletes half of the original 32 components in the coordinates $\theta^\mu$.

We first analyze our field equations, starting with that of the $\mu$-field

$$\rho^{ijk}_{(+)}\lambda^{lmn}_{(+)} = 0 \ . \tag{4.10}$$

There are two solutions for this field equation: $\rho^{ijk}_{(+)} = 0$, and/or $\lambda^{ijk}_{(+)} = 0$. However, the former leads to the trivial fermionic transformation in (4.8a), to be excluded as a trivial option, so we concentrate on the latter solution:

$$\lambda^{ijk}_{(+)} = 0 \ . \tag{4.11}$$

The $b$-field equation immediately yields $\partial_i \lambda = 0$, i.e., $\lambda = \text{const.} \equiv C$, while the $\nu$-field equation $\lambda \rho^{ijk}_{(+)} = 0$ fixes this constant $C$ to be zero:

$$\lambda = 0 \ . \tag{4.12}$$

Now the $\rho$-field equation

$$\mu^{(-)(-)}_{ijklmn} \lambda^{lmn}_{(+)} + \nu^{(-)}_{ijk} \lambda = 0 \ , \tag{4.13}$$

automatically holds under (4.11) and (4.12). Similarly, the $a$-field equation

$$3\partial_k \lambda^{ijk}_{(+)} - \tfrac{1}{24}\epsilon^{ijk_1\cdots k_4} F_{k_1\cdots k_4} \lambda = 0 \ , \tag{4.14}$$

is also satisfied by (4.11) and (4.12). The $\lambda$-field equation reads

$$\tfrac{2}{6!}\epsilon^{i_1\cdots i_6} g_{i_1\cdots i_6} + e + \nu^{(-)}_{ijk} \rho^{ijk}_{(+)} = 0 \ . \tag{4.15}$$

The metric $g_{ij}$-field equation is easy to see, because under (4.12) the only contribution is from $\mathcal{L}_{\text{det}}$, whose general variation is[9]

$$\delta\mathcal{L}_{\text{det}} = -\tfrac{1}{2}\left[\det\left(g_{kl} - \Pi_k{}^c \Pi_{lc}\right)\right]^{-1/2} \left(g_{ij} - \Pi_i{}^a \Pi_{ja}\right) \delta(g^{ij} - \Pi^{ib}\Pi^j{}_b) \ . \tag{4.16}$$

Thus the metric equation implies the embedding condition

$$g_{ij} = \Pi_i{}^a \Pi_{ja} \ . \tag{4.17}$$

Under this condition, the $Z^M$-field equation is also satisfied, because under (4.17) the only possible contribution to this equation from $\mathcal{L}_{\lambda f}$ or $\mathcal{L}_{\lambda g}$ again vanishes under (4.11), e.g., $\delta\mathcal{L}_{\lambda f}/\delta Z^M = \lambda^{ijk}_{(+)}(\delta f_{ijk}/\delta Z^M) = 0$. The only remaining field equation is that of $\lambda^{[3]}$:

$$f^{(-)}_{ijk} + \mu^{(-)(-)}_{lmnijk} \rho^{lmn}_{(+)} = 0 \ . \tag{4.18}$$

This field equation does *not* necessarily implies the self-duality $f^{(-)} = 0$, but it is *allowed* as a sufficient condition:

$$f^{(-)}_{ijk} = 0 \ , \quad \mu^{(-)(-)}_{ijklmn} = 0 \ . \tag{4.19}$$

This feature that the self-duality of $f$ is not forced by a field equation, but is allowed only as a sufficient condition, is expected from the general argument of M-theory [7], and also similar to ref. [11]. A more generalized self-duality: $f^{(-)}_{ijk} = c f_{[i|}{}^{mn} f_{mn}{}^l f_{l|jk]}$ in [11] can be also embedded in (4.18) by $\rho_{(+)}^{lmn} = f_{(+)}^{lmn}$ and $\mu_{ijk}{}^{lmn} = (c/6) f_{[i}{}^{[lm} f^{n]}{}_{jk]}$ with the appropriate duality projection. From (4.8e) it is also clear that the solution (4.19) is not covariant under our fermionic symmetry, similarly to [11]. This is also natural, because fermionic symmetries can be generally fixed, in such a way that unwanted states are eliminated.

---

[9]Note that the undesirable singularity at $g_{ij} - \Pi_i{}^a \Pi_{ja} = 0$ in $\delta\mathcal{L}_{\text{det}}$ can be easily avoided by an alternative lagrangian: $\mathcal{L}'_{\text{det}} \equiv e^{1-m}\left[\det\left(g_{ij} - \Pi_i{}^a \Pi_{ja}\right)\right]^{m/2}$ for a real number $m > 2$.

We now study the fermionic invariance of our total action. By the help of the useful relationship [11]

$$\delta_\zeta f_{ijk} = -(\delta_\zeta E^B) F_{Bijk} = -\tfrac{1}{4}(\overline{\kappa}_- \gamma_{[ij} \Pi_{k]}) = -i\tfrac{1}{4}(\overline{\zeta}_+ \gamma_{[3]} \gamma_{[ij} \Pi_{k]}) \rho_{(+)}^{[3]} \ , \qquad (4.20\text{a})$$

$$\delta_\zeta g_{i_1 \cdots i_6} = -i\tfrac{1}{240}(\overline{\kappa}_- \gamma_{[i_1 \cdots i_5} \Pi_{i_6]}) + \tfrac{1}{24}(\overline{\kappa}_- \gamma_{[i_1 i_2} \Pi_{i_3}) f_{i_4 i_5 i_6]} \ , \qquad (4.20\text{b})$$

we easily see that

$$\begin{aligned}
0 \stackrel{?}{=} \delta_\zeta \mathcal{L} &= \delta_\zeta (\mathcal{L}_{\lambda f} + \mathcal{L}_{\lambda g} + \mathcal{L}_{\mu \rho \lambda} + \mathcal{L}_{\nu \rho \lambda}) \\
&= + \lambda_{(+)}^{ijk} (-\tfrac{3}{2})(\overline{\kappa}_- \gamma_{ij} \Pi_k) + \lambda \left[ ie(\overline{\kappa}_- \gamma_7 \gamma^i \Pi_i) + \tfrac{1}{6} e(\overline{\kappa}_- \gamma_7 \gamma^{[3]} \Pi_i) f_{[3]} - ie(\overline{\kappa}_- \gamma^i \Pi_i) \right] \\
&\quad - i\tfrac{1}{4}(\overline{\zeta}_+ \gamma_{[3]} \gamma^i \gamma_{[3]'} \Pi_i) \rho_{(+)}^{[3]} \lambda_{(+)}^{[3]'} - i\tfrac{1}{6} e\lambda (\overline{\zeta}_+ \gamma_{[3]} \gamma^{[3]'} \Pi_i) f_{[3]'} \rho_{(+)}^{[3]} \\
&= + \tfrac{1}{4}(\overline{\kappa}_- \gamma^k \gamma_{[3]} \Pi_k) \lambda_{(+)}^{[3]} + i\tfrac{1}{6} e\lambda (\overline{\zeta}_+ \gamma_{[3]} \gamma^{[3]'} \Pi_i) \rho_{(+)}^{[3]} f_{[3]'} \\
&\quad - i\tfrac{1}{4}(\overline{\zeta}_+ \gamma_{[3]} \gamma^i \gamma_{[3]'} \Pi_i) \rho_{(+)}^{[3]} \lambda_{(+)}^{[3]'} - i\tfrac{1}{6} e\lambda (\overline{\zeta}_+ \gamma_{[3]} \gamma^{[3]'} \Pi_i) \rho_{(+)}^{[3]} f_{[3]'} \\
&= + i\tfrac{1}{4}(\overline{\zeta}_+ \gamma_{[3]} \gamma^i \gamma_{[3]'} \Pi_i) \rho_{(+)}^{[3]} \lambda_{(+)}^{[3]'} - i\tfrac{1}{4}(\overline{\zeta}_+ \gamma_{[3]} \gamma^i \gamma_{[3]'} \Pi_i) \rho_{(+)}^{[3]} \lambda_{(+)}^{[3]'} = 0 \ , \qquad (4.21)
\end{aligned}$$

based on relations such as $\gamma_7 \gamma_{[3]} \lambda_{(+)}^{[3]} = -\gamma_{[3]} \lambda_{(+)}^{[3]}$, $\overline{\kappa}_- \gamma_7 = +\overline{\kappa}_-$, etc.

In our formulation, we have no $\sigma$-model type kinetic term. This situation is similar to ref. [11], but the system allows a generalized self-duality for $f_{ijk}$, as a special case. This is natural in the super M-5-brane formulation, in the sense that the physical field is now $f_{ijk}$, instead of the $\sigma$-model coordinates $Z^M$. It is interesting that our auxiliary field $\mu_{ijklmn}^{(-)(-)}$ has the index symmetries similar to the 11D superspace auxiliary superfield $\Lambda^{a_1 \cdots a_4 b_1 \cdots b_4}$.

## 5. Concluding Remarks

In this paper, we have presented a very simple duality-symmetric local lagrangian formulation that utilizes only one tensor multiplier field $\Lambda^{[4][4]'}$. Compared with the recent paper on a similar subject [13], our formulation has simpler constraint lagrangians whose structure is essentially the same as that proposed by Siegel [18]. The corresponding superspace formulation with the manifest duality relation between $F_{[4]}$ and $G_{[7]}$ is straightforward. Also presented is a new super M-5-brane action, allowing the self-duality for the field strength $f_{ijk}$, formulated on our 11D duality-symmetric supergravity backgrounds.

We saw that our new super M-5-brane action has couplings more natural than the formulation in [11], in the sense that the embedding condition of the 6D metric in terms of the pull-back comes out as a field equation. Our lagrangian is also simpler than the formulation using the unit vector with scalar field [14][9]. It is also interesting to see if the dual version for $N=1$ supergravity in 10D [25] can be re-obtained by performing double-dimensional reductions [26] into superstrings.

Note that our invariant lagrangian does *not* force the field strength $f_{ijk}$ to be self-dual as a *necessary* condition, but instead, the self-duality is allowed as a *sufficient* condition. Therefore our lagrangian bypasses the obstruction for constructing an invariant lagrangian for a chiral two-form in 6D [7][27]. This obstruction was from various considerations leading to the conclusion that a chiral two-form in 6D can not have an invariant modular form that is needed for an invariant lagrangian [7][27]. In other words, since our field strength $f_{ijk}$ is not necessarily a self-dual field, we can construct an invariant lagrangian. This feature is similar to that in [11].

The method we used in 11D resembles that in [18] with a constraint lagrangian bilinear in the constraint and linear in the multiplier field, with eight indices in our case. Such a system prevents the multiplier field from propagating. A similar method is also used in our 6D super M-5-brane action. We have also seen that the existence of the $F_{[4]}^2$ and $G_{[7]}^2$-terms are crucial for the multiplier field to be gauged away by the extra symmetry (2.20) in the standard manner [18]. For this reason we should avoid the singular cases $\beta = 0$, $\beta = (24)^{-1}$. Despite of the eight-index auxiliary field $\Lambda^{[4][4]'}$, our action is much simpler than that in [13] utilizing a scalar field $a(x)$ with its gradient $v_\mu \equiv [(\partial_\nu a)^2]^{-1/2} \partial_\mu a$ [14] whose *non-invariance* under supersymmetry[10] makes the computation more involved.

Our 11D superspace BIs are satisfied, only if the new $\widehat{\mathcal{F}}$-dependent terms vanish. In other words, these 'on-shell vanishing' $\widehat{\mathcal{F}}$-terms do not satisfy the BIs, in contrast with the usual off-shell formulation in superspace supergravity, where all the auxiliary-dependent terms also satisfy the BIs. In this sense, our superspace constraints are on-shell equivalent to those in ref. [1], like the formulation with scalar auxiliary superfields in [13].

As long as $\beta \neq 0$, $\beta \neq (24)^{-1}$ for the possible $\eta$-symmetry, there are always kinetic terms both for $A_{[3]}$ and $B_{[6]}$, and conjugate momenta for these fields exist, even though only half of the total degrees of freedom are counted as physical ones by the duality by the $\Lambda$-field equation. In other words, only the cases $\beta \neq 0$, $\beta \neq (24)^{-1}$ seem to allow simple quantization. This feature is more elucidated in our formulation than that with scalar field [14][13], and makes our formulation practically more useful.

Thanks to the simplicity of the system, our formulation provides a good working ground for studying various aspects of M-theory, such as D-brane couplings with a two-form field strength, more unified super M-5-brane couplings with self-dual three-form field strength, non-perturbative aspects, double-dimensional reduction to supermembrane [2][26], or relationships with the dual formulation in 10D [25].

Special acknowledgements are for M. Cederwall, N. Berkovits, S.J. Gates, Jr., B.E.W. Nilsson, and W. Siegel for important communications.

---

[10]Note that the unit vector $v_\mu$ in [14][13] is *not* invariant under supersymmetry due to the metric tensor involved in the scalar product $(\partial_\nu a)^2 \equiv g^{\mu\nu}(\partial_\mu a)(\partial_\nu a)$.